\documentclass[prd,twocolumn,amsmath,amssymb,floatfix,superscriptaddress]{revtex4-1}

\usepackage{graphicx}
\usepackage{amssymb}
\usepackage{amsmath}
\usepackage{hyperref}

\usepackage{color}

\DeclareMathAlphabet\mathbfcal{OMS}{cmsy}{b}{n}
\definecolor{darkgreen}{cmyk}{0.85,0.2,1.00,0.35} 
\definecolor{purple}{cmyk}{0.5,1.0,0,0}

\newcommand{\stucky}{St\"{u}ckelberg}
\newcommand{\tr}[1]{[#1]}

\newcommand{\Mpl}{M_{\rm pl}}
\newcommand{\scalar}{\delta_{v0}^{\rm K}}

\newcommand{\dd}{\mathrm{d}}
\newcommand{\reff}[1]{(\ref{#1})}

\newcommand{\ul}[3]{#1^{#2}_{ \hphantom{#2}\! #3}}
\newcommand{\lu}[3]{#1_{#2}^{ \hphantom{#2}\! #3}}
\newcommand{\fid}{\Sigma}
\newcommand{\bgamma}{\boldsymbol{\gamma}}

\newcommand{\bargamma}{\bar{\gamma}}
\newcommand{\aM}{V} 
\newcommand{\OurG}{\chi}
\newcommand{\bOurG}{{\boldsymbol{\chi}}}
\def\barray{\begin{array}}
\def\earray{\end{array}}
\def\be{\begin{equation}}
\def\ee{\end{equation}}
\def\ben{\begin{equation} \nonumber}
\def\een{\end{equation}}
\def\ban{\begin{eqnarray*}}
\def\ean{\end{eqnarray*}}
\def\ba{\begin{eqnarray}}
\def\ea{\end{eqnarray}}

\def\({\left(}
\def\){\right)}

\begin{document}

\title{Self-accelerating Massive Gravity: Covariant Perturbation Theory}
\author{Pavel Motloch}
\affiliation{Kavli Institute for Cosmological Physics, Department of Physics, University of Chicago, Chicago, Illinois 60637, U.S.A}
\author{Wayne Hu}
\affiliation{Kavli Institute for Cosmological Physics, Department of Astronomy \& Astrophysics,  Enrico Fermi Institute, University of Chicago, Chicago, Illinois 60637, U.S.A}
\begin{abstract}
We undertake a complete and covariant treatment for the quadratic Lagrangian of all of the degrees of freedom
of massive gravity with a fixed flat fiducial metric  for arbitrary massive gravity parameters around any isotropic self-accelerating background solution.  Generically,  3 out of 4 \stucky\ degrees of freedom propagate
in addition to the usual 2 tensor degrees of freedom of general relativity.   The complete
kinetic structure typically is only revealed at an order in the graviton mass that is  equivalently 
to retaining curvature terms in a locally flat expansion.   These results resolve several apparent
discrepancies in the literature where zero degrees of freedom propagate in either special
cases or approximate treatments as well as decoupling limit analyses which attempt
to count longitudinal degrees of freedom.  
\end{abstract}

\maketitle
\section{Introduction}

The theory of massive gravity with a second static flat fiducial  metric \cite{Gabadadze:2009ja,deRham:2009rm,deRham:2010ik,deRham:2010kj}  possesses solutions that accelerate the
 cosmological expansion in the
absence of a true cosmological constant  \cite{deRham:2010tw,Koyama:2011xz,Koyama:2011yg,Nieuwenhuizen:2011sq,Berezhiani:2011mt,D'Amico:2011jj,Gumrukcuoglu:2011ew,Gratia:2012wt,Kobayashi:2012fz,Volkov:2012cf,Volkov:2012zb}. 
Because the second metric is non-dynamical, this theory of massive gravity breaks
diffeomorphism invariance.   While covariance can be restored with the \stucky\ trick,
 for a homogeneous and isotropic spacetime background, the
\stucky\ fields must be inhomogeneous to accommodate the two metrics.   Moreover,
except for a special class of  open universe solutions \cite{Gumrukcuoglu:2011ew}, 
there is no coordinate system where both the spacetime and fiducial metrics can 
be made simultaneously diagonal, homogeneous and isotropic \cite{D'Amico:2011jj} even though the spacetime metric itself can accommodate self-accelerating solutions with Friedman-Robertson-Walker
backgrounds for any desired matter content or curvature \cite{Gratia:2012wt}.

Inhomogeneity in the \stucky\ fields or equivalently the relationship between the
spacetime and Minkowski metrics  causes both technical
and theoretical challenges for understanding the self-accelerating solutions.
In the special open universe  case where the metrics themselves are simultaneously homogeneous
and isotropic in the same coordinates, standard analyses apply. There the massive gravity sector is strongly
coupled and propagates no degrees of freedom in the quadratic Lagrangian and possesses
an instability at higher order \cite{Gumrukcuoglu:2011zh}.  Furthermore
the solutions can evolve to a coordinate singularity which cannot be resolved by 
charts with overlapping domains of validity \cite{Gratia:2013gka}.  

For the more generic case, technical difficulties of incorporating an inhomogeneous
fiducial metric background, which breaks translation invariance, has hitherto prevented a full analysis.    Results for the longitudinal or isotropic modes with the exact background obtained in Ref.~\cite{Wyman:2012iw} showed  one propagating degree of freedom in the quadratic Lagrangian, as
might be expected from the 4 \stucky\ fields, the Boulware-Deser ghost free construction
and isotropy indicating that the behavior of the open universe solution is not generic.
However this degree of freedom obeys unusual but stable first order dynamics with no quadratic
coupling to other fields \cite{Wyman:2012iw} and an unbounded Hamiltonian 
\cite{Khosravi:2013axa}.    Furthermore these results 
seem to contradict analyses that used  the decoupling limit
\cite{Koyama:2011wx} or locally flat approximations  \cite{D'Amico:2012pi}
which found that generically two and zero isotropic modes propagate respectively in
the quadratic Lagrangian.   
Indeed in the locally flat approximation, no anisotropic modes propagated either.

It is the aim of this paper to resolve these issues and present a complete covariant
treatment of the quadratic Lagrangian for the massive gravity degrees of freedom
around any  isotropic  self-accelerating background.
We begin in \S \ref{sec:theory} with a brief review of the massive gravity theory to establish 
notation.   In \S \ref{sec:decoupling} we reanalyze the isotropic modes for the class
of solutions considered in the existing literature and show that inconsistencies in the
counting of degrees of freedom are resolved by a complete analysis at the level of first order
curvature corrections to the locally flat approximation and a consistent treatment of
gauge degrees of freedom.  Since the full dynamics of the modes involves spacetime curvature,  in \S \ref{sec:vectors} we provide a general, covariant 
treatment of {\it all} massive gravity degrees of freedom for {\em any} isotropic background solution
on the self-accelerating branch for the {\em entire} class of massive gravity parameters.   We discuss these results in \S \ref{sec:discussion}.

\section{Massive Gravity}
\label{sec:theory}

The Boulware-Deser ghost free theory of massive gravity adds a mass term to the Einstein-Hilbert Lagrangian density
 \cite{deRham:2010kj}
\begin{align}
\label{drgt}
\mathcal{L}^{\rm (M G)} &=-\frac{m^2 M_{\rm pl}^2}{2}\sqrt{-g}\sum_{k=0}^4 \frac{\beta_k}{k!} F_k\left(\bgamma\right),
\end{align}
where $M_{\rm pl}=(8\pi G)^{-1}$ is the reduced Planck mass and the $F_k$ terms are
functions of the matrix $\bgamma$
\begin{align}
F_0(\bgamma) & = 1, \nonumber\\
F_1(\bgamma) & = \tr{\bgamma}, \nonumber\\
F_2(\bgamma) & =  \tr{\bgamma}^2 - \tr{\bgamma^2} , \\
F_3(\bgamma) & =\tr{\bgamma}^3 - 3 \tr{\bgamma} \tr{\bgamma^2} + 2 \tr{\bgamma^3} , \nonumber\\
F_4(\bgamma) &= \tr{\bgamma}^4 - 6 \tr{\bgamma}^2 \tr{\bgamma^2} + 3 \tr{\bgamma^2}^2 + 8 \tr{\bgamma} \tr{\bgamma^3}  
- 6 \tr{\bgamma^4} ,
\nonumber
\end{align}
where $[\,]$ denotes the trace of the enclosed matrix.  
The parameters of the theory are
$m$, the graviton mass, and $\beta_k$.  Not all of the latter parameters are independent since
\begin{align}
\beta_0 &= -12 (1+ 2\alpha_3+2\alpha_4), \nonumber\\
\beta_1 &= 6(1 + 3 \alpha_3 + 4\alpha_4),\nonumber\\
\beta_2 &= -2(1+ 6 \alpha_3+12\alpha_4 ), \\
\beta_3 &= 6(\alpha_3+ 4\alpha_4), \nonumber\\
\beta_4 &= -24 \alpha_4,\nonumber
\end{align} 
leaving two remaining independent parameters $\{\alpha_3,\alpha_4\}$.

The presence of the matrix $\bgamma$ breaks diffeomorphism invariance since it is 
constructed from the square root of the product of the inverse spacetime metric $g^{\mu\nu}$ and a flat fiducial metric $\fid_{\mu\nu}$
\begin{equation}
\ul{\gamma}{\mu}{\alpha} \ul{\gamma}{\alpha}{\nu} = g^{\mu\alpha}\fid_{\alpha\nu},
\end{equation}
singling out a specific coordinate choice, called unitary gauge where $\fid_{\mu\nu}=\eta_{\mu\nu}$, the Minkowski metric.   Nonetheless,  diffeomorphism invariance can be restored by the \stucky\ trick of using the coordinates of unitary gauge as 
 auxiliary fields $\phi^a$ to  express the fiducial metric in an arbitrary coordinate system
 \begin{equation}
\fid_{\mu\nu} = \partial_\mu\phi^a \partial_\nu \phi^b \eta_{a b}.
\label{eqn:stucky}
\end{equation}
It is important to note that 
the \stucky\ fields $\phi^a$ transform as spacetime scalars and form a Lorentz 
vector only in the internal Minkowski space. 
Beyond the leading order, locally flat approximation to the spacetime metric, \stucky\
indices should not be conflated with spacetime indices \cite{deRham:2011rn}.   We
shall see in \S \ref{sec:vectors} how to construct spacetime vectors out of \stucky\ components.
 Throughout, Greek indices denote the spacetime and are lowered and raised with $g_{\mu\nu}$ and its inverse; Latin indices likewise by the Minkowski metric $\eta_{ab}$.

\section{Isotropic Modes}
\label{sec:decoupling}

In this section,
we resolve discrepancies in the literature for the dynamics of spherically symmetric \stucky\ perturbations around certain self-accelerating vacuum solutions \cite{Koyama:2011yg}, first analyzed in the decoupling limit \cite{Koyama:2011wx}, then in a locally flat limit \cite{D'Amico:2012pi}, and finally in the
exact background \cite{Wyman:2012iw,Khosravi:2013axa} with contradictory results.
In \S \ref{sec:conformalkinetic} we treat the quadratic Lagrangian consistently to
 leading order in  curvature corrections, ${\cal O}(m^2)$, and show that kinetic terms for a single  longitudinal or isotropic mode only 
 arise at this order, which is then consistent with exact results.   In \S \ref{sec:scaling} we discuss the problem of equating a locally
 flat expansion with the decoupling limit around a Minkowski background.   In \S \ref{sec:gauge}, we show that miscounting of degrees of freedom can also arise from gauge
 fixing in the Lagrangian.

\subsection{Kinetic Terms and Curvature}
\label{sec:conformalkinetic}

We focus here only on the specific case of certain solutions  for the $\alpha_3 = \alpha_4 = 0$
model   \cite{Koyama:2011yg} as these suffice to show our main points and have been the
most analyzed in the literature.   {We consider the general case in \S \ref{sec:covariantform}.}  With this choice, the unitary gauge solution is described by the metric
\ba
	g_{\mu\nu} \dd x^\mu \dd x^\nu &=& - {\cal C}(R) \dd T^2 + 2 {\cal D}(R) \dd T \dd R+ {\cal A}(R) \dd R^2\nonumber\\
		& &+ {\cal B}(R)\(\dd \theta^2 + \sin \theta^2 \dd \phi^2\) ,
\ea
where
\ba
	{\cal A}(R) &=& \frac{4 C^2}{9}\left(1+ v^2 + \frac{m^2 R^2}{9}  \right),\nonumber\\
	{\cal B}(R) &=& \frac{4}{9} R^2,\nonumber\\
	{\cal C}(R) &=& \frac{4 C^2}{9}\left(1 - \frac{m^2 R^2}{9}\right),\nonumber\\
	{\cal D}(R) &=& \frac{4 C^2 }{9} \frac {m R}{3} \sqrt{  v^2 + \frac{m^2 R^2}{9} },
	\ea 
	with \footnote{Our parameters $v, C$ are related to those used in \cite{Koyama:2011yg} as $Q_0 = {3v}/{4}$, $\Delta_0 = {16 C^2}/{81}$, $\gamma = v^2$.}
	\ba	
	v^2 &=& \frac{1}{C^2} - 1 .
\ea
Here $0 < C \le 1$ is an integration constant in the solutions.  From this
exact expression we would like to focus on a locally flat patch where $m R \rightarrow 0$.
There is a subtlety in taking this limit associated with the parameter $v$.    The metric component
\begin{equation}
{\cal D}(R) =  \frac{4 C^2 }{9} \frac {m R}{3}  v \left[ 1+ \frac{m^2 R^2}{18v^2} +
{\cal O}\left( \frac{ m^4 R^4}{v^4}\right) \right].
\end{equation}
Hence the validity of the expansion is confined to a radius where $m R \ll v$ 
rather than $m R \ll 1$.  For the special case of $v=0$ ($C=1$), this domain
of validity shrinks to $R=0$ and instead 
\begin{equation}
{\cal D}(R) = \frac{4}{9} \left( \frac{m R}{3}\right)^2, \qquad (v=0).
\end{equation}    The result is 
an apparent discontinuity in the limit $v \rightarrow 0$ of the expansion.  We shall
see that this limit is the case where the background solution has no \stucky\ vector component.

Next when considering fluctuations around this background solution it is more convenient to 
choose a different gauge where  the background metric is described near the origin $r=t=0$ to order ${\cal O}(m^2)$
by the conformal form  \cite{Koyama:2011wx} 
\be
	\bar{g}_{\mu\nu} =  \left[ 1 - \frac{m^2(r^2-t^2)}{8}\right] \eta_{\mu\nu} + {\cal O}(m^3),
	\label{conformal}
\ee
where 
 $\eta_{\mu\nu}$ is the Minkowski metric in spherical coordinates
\be
	\eta_{\mu\nu} \dd x^\mu \dd x^\nu = - \dd t^2 + \dd r^2 + r^2 \(\dd \theta^2 + \sin \theta^2 \dd \phi^2\).
\ee
The unitary gauge coordinates $(T, R)$ can be expressed in conformal coordinates $(t, r)$ as
\ba
	\label{backgrT}
	 T &=& \frac{3 t}{2C }\left[ 1 + \frac{m^2 (t^2+3 r^2)}{48} \right]
	 + \frac{ mr^2 }{8} (3v  +\scalar  m r)\nonumber\\
	 && +{\cal O}(m^3),\nonumber \\ 
	 %
	 \label{backgrR}
    R &=& \frac{3 r}{2} \left[ 1 - \frac{m^2(r^2-t^2)}{16}\right] + {\cal O}(m^3) .
\ea
Here, the Kronecker delta
\begin{equation}
\scalar = \begin{cases}
1 & v=0 \\
0 & v\ne 0\\
\end{cases}
\end{equation}
accounts for the discontinuity at $v=0$ in the expansion of ${\cal D}(R)$ above.  Note that
the discontinuity appears only at ${\cal O}(m^2)$ and hence was omitted in
Ref.~\cite{Koyama:2011wx}.  For $v\ne 0$, unitary time $T$ contains terms that are  ${\cal O}(m)$.

The background \stucky\ fields are then the unitary gauge coordinates $(T,R)$ of the
solution \cite{Koyama:2011yg}
 expressed
in conformal coordinates $(t,r)$
\ba
	\bar \phi^0 &=& T,\nonumber\\
	\bar \phi^i &=& R \frac{x^i}{r} .
	\label{stuckyback}
\ea
 Note that in the special case that $v=0$,  the
difference in the unitary and conformal coordinates can be derived from a Lorentz scalar quantity ignoring curvature corrections
\begin{equation}
\bar\phi_\mu - x_\mu =\frac{1}{4} \partial_{\mu}( r^2- t^2) + {\cal O}(m^2).
\end{equation}
This $v=0$ case is said to have no vector \stucky\ field in the background.  Note that  the
\stucky\ index, which is always raised an lowered by the Minkowski metric, cannot be treated
as a spacetime index at ${\cal O}(m^2)$.  In Ref.~\cite{Koyama:2011wx}, the background \stucky\ were truncated already before ${\cal O}(m^2)$.
However we shall see that since the dynamics of perturbations enter at $O(m^2)$, this is not sufficient.

To see this consider spherically symmetric fluctuations in the \stucky\ fields
\ba
	\delta \phi^0 &\equiv& -a_t(t,r),\nonumber\\
	\delta \phi^i &\equiv& a_r(t,r) \frac{x^i}{r},
\ea
where we keep the definition that \stucky\ indices are raised and lowered by the Minkowski
metric at the expense of $a_\mu$ not forming a spacetime vector at $O(m^2)$.   We
correct this notational abuse in \S \ref{sec:vectors}.   Expanding the Lagrangian (\ref{drgt}) to 
quadratic order in the \stucky\ fluctuations, we obtain 
\ba\label{OurKoyamaL}
	{\mathcal{L}}_2 &=& \Mpl^2 m^2  \frac{(m r)^2 \sin\theta}{4(1+C)}\Big[ 
	 \Big( \frac{4 Cv }{mr }  +2 \scalar\Big)  r a_r a_t' + 3 C   a_r^2 \nonumber\\
	&&  + 2  r a_t \dot a_r + {\cal O}(m) \Big] .
\ea
Here and in the following, we equate Lagrangians which are equal up to
total derivative terms. The primes denote derivatives with respect to the radial coordinate $r$, dots with respect to $t$.

Note that we do not consider mixing with metric perturbations here.  Unlike in the Minkowski background, the longitudinal or isotopic mode gains a kinetic term 
from curvature corrections rather than demixing with the metric fluctuations
\cite{Mirbabayi:2011aa} as can explicitly be shown given that the self-accelerating solution is
exact for isotropic metrics, perturbed or not \cite{Wyman:2012iw}. We shall 
return to this topic in \S \ref{sec:vectors} where anisotropic modes and their mixing are 
considered.

    Aside from the overall factor of  $\Mpl^2 m^2$ from Eq.~(\ref{drgt}), the leading order
terms in Eq.~(\ref{OurKoyamaL}) come in at ${\cal O}(m^2)$ if there is no vector background and ${\cal O}(m)$ with
a vector background.   Furthermore even in the latter case the ${\cal O}(m)$ terms carry no
time derivatives and thus propagate no degrees of freedom.    This explains the
result of Ref.~\cite{D'Amico:2012pi}, where all terms of order ${\cal O}(m^2)$ were omitted.
The dynamical term from $\dot a_r$ only enters in at ${\cal O}(m^2)$ whereas $a_t$
is non-dynamical.    The first order structure of this Lagrangian implies that $a_r$ obeys
a first order equation of motion that is  independent of $a_t$ whereas $a_t$ obeys a first order
equation that depends on $a_r$.   This result is in accordance with the exact background
\cite{Wyman:2012iw}.

This set of equations of motion do not combine into the usual wave equation for
one degree of freedom.   Nonetheless a Hamiltonian analysis shows that $(a_r,a_t)$
form a single degree of freedom due to the presence of constraints. 
In particular the field momenta 
\ba
	p_{a_t} &=& \frac{\partial \mathcal{L}_2}{\partial \dot a_t} = 0,\nonumber\\
	p_{a_r} &=& \frac{\partial \mathcal{L}_2}{\partial \dot a_r} = \frac{\Mpl^2 m^4 r^3\sin \theta}{2(1+C)}a_t ,
\ea
cannot be inverted to express velocities $\dot a_t, \dot a_r$ in terms of momenta, which indicates the presence of two primary constraints
\ba
	\phi_1 &=& p_{a_t},\nonumber\\
	\phi_2 &=& p_{a_r} - \frac{ \Mpl^2 m^4 r^3 \sin \theta}{2(1+C)}a_t\label{phi2}.
\ea
To determine whether $(\phi_1,\phi_2)$ exhausts all of the constraints we define
 the total Hamiltonian
%
\ba
	\mathcal{H}_T &=& - \frac{ \Mpl^2 m^4 r^2\sin \theta}{4(1+C)}\Big[
	 \Big( \frac{4 Cv}{mr }  +2 \scalar\Big)  r a_r a_t'+ 3 C  a_r^2\Big] \nonumber \\
	& &+ u_1(t,r) \phi_1 + u_2 (t,r) \phi_2,
\ea
where $u_i$ are (at the moment) auxiliary variables.   Using Poisson brackets and
the Hamilton equation, we see that
\be
	 \dot \phi_i = \{ \phi_i, \mathcal{H}_T \} =0
\ee
uniquely determine $u_i$ and we thus conclude there are no additional constraints in the system.

Because  $\{ \phi_1, \phi_2\}$ 
is nonzero, both primary constraints are second class and together remove one of the two degrees of freedom from the problem. This leaves us with a single propagating degree of freedom. 

This result is in full agreement with that of the exact theory \cite{Wyman:2012iw,Khosravi:2013axa}, which also found one propagating degree of freedom for all values of $C$. On the constrained surface $\phi_i = 0$ we can utilize the constraint \reff{phi2} and rewrite the Hamiltonian entirely in terms of $a_r, p_{a_r}$:
\ba
	\mathcal{H}\! &=&\! \frac{2 C v}{m r} \!\!\left( \frac{2}{r} a_r+ a_r' \!\right)  \!p_{a_r} \! + \scalar \!\!\left( \frac{3}{r} a_r+ a_r' \!\right)  \!p_{a_r} \!
	\nonumber\\
	 && - \frac{3 C \Mpl^2 m^4 r^2 \!\sin\theta}{4(1+C)}{} a_r^2.
\ea

The Hamiltonian is linear in the now unbounded $p_{a_r}$,  which means the Hamiltonian is unbounded from below. This is also in agreement with the results of the full theory.
We concluded that an expansion to the level of curvature corrections in the locally
flat limit is sufficient to recover the \stucky\ dynamics.

\subsection{Scaling vs.\ Decoupling}
\label{sec:scaling}

The analysis in Ref.~\cite{Koyama:2011wx} is based on a scaling limit with $m$ motivated
by the decoupling of scalars, vectors and tensors or helicity states of the graviton around a Minkowski background.   We 
show here why this limit is misleading for the self-accelerating background 
given a lack of decoupling.   Naive use of this scaling limit would erroneously imply zero
degrees of freedom rather than one.

Around a Minkowski background, scaling the \stucky\ fluctuations according to 
\ba
	\label{NaiveSV}
	a_\mu &=& \frac{ m A_\mu + \partial_\mu \pi}{\Mpl m^2}
\ea
leads to the so-called decoupling limit where $m \rightarrow 0$ at fixed $\Mpl m^2$. 
In this limit 
 $A_\mu$ is a free vector field with a canonical  Maxwell Lagrangian
and $\pi$  is a scalar which gets a canonical kinetic term once demixed from the tensor metric fluctuation \cite{ArkaniHamed:2002sp}.   
In the spherically symmetric configuration studied in this section, this would lead to one scalar
or helicity-0 mode since the Maxwell Lagrangian propagates only transverse degrees of freedom.  In this sense $\pi$ is an additional \stucky\ field that restores $U(1)$ symmetry
to the vector 
by separating out its longitudinal component.  We shall return to this point
in the next section.

Rewriting the \stucky\ fluctuations in the form Eq.~(\ref{NaiveSV}) of course cannot
change the dynamics or the number of propagating degrees of freedom.  The problem
is that the motivation for this scaling disappears around the self-accelerating background
where kinetic terms only come in at ${\cal O}(m^2)$.     We shall therefore refer to the 
decomposition of Eq.~(\ref{NaiveSV}) as the Minkowski scaling limit rather than the decoupling limit.

With this scaling  we can write the $v\ne 0$ Lagrangian of Eq.~(\ref{OurKoyamaL}) as
an  ${\cal O}(m^0)$ term 
\be
	\label{Lagr2Koyama}
	\mathcal{L}_2^{(0)} =
	\frac{r^2 \sin \theta}{1 + C }\left[ C v \pi' (A_t' - \dot A_r ) 
		+ \frac{3}{4} (C \pi'^2 -   \dot \pi ^2) \right]
\ee
plus terms that appear to be higher order
\ba
   \mathcal{L}_2 &=&  
   \mathcal{L}_2^{(0)} + 
   \frac{r^2 \sin \theta}{4\( 1 + C \)}\Big\{  m \Big[  2 r \dot A_r \dot \pi +4 C v A_r A_t'        \nonumber \\   
   & &  
      + 6 C A_r \pi' + 2 r A_t \dot \pi' + 2 \scalar r \pi' ( A_t' - \dot A_r)\Big]
		\nonumber\\
	& &	
		+ m^2 ( 3 C A_r^2 + 2 r A_t \dot A_r + 2 \scalar r A_r A_t')
		\Big\}.  
\ea
 Note that unlike the  decoupling limit, $A_\mu$ does not possess a Maxwell term nor does it decouple from $\pi$.  
Ref.~\cite{Koyama:2011wx} kept only the $\mathcal{L}_2^{(0)}$ term [see 
their Eqs.~(5.8), (5.9) and (5.26)] based on the assumption that taking the 
$m\rightarrow 0$ with the Minkowski scaling was self-consistent.   However, we know that ($A_t,A_r, \pi$) together form a single degree of freedom. Dropping any interaction between these fields by simply assuming that
$A_\mu$ should scale differently with $m$ is thus dangerous.

Indeed the Hamiltonian analysis shows 
$\mathcal{L}^{(0)}_2$ 
propagates no degrees of freedom for $v \neq 0$. The primary constraints are
\ba
	\phi_1 &=& p_{A_t},\nonumber\\
	\phi_2 &=& p_{A_r} + \frac{C v r^2 \sin \theta }{1+C}\pi' ,
\ea
while we are able to express the velocity $\dot \pi$ in terms of the momentum $p_\pi$.

Time evolution of these two constraints by calculating their Poisson brackets with the total Hamiltonian provides two secondary constraints
\ba
	\phi_3 &\sim& \(r^2 \pi'\)',\nonumber\\
	\phi_4 &\sim& r^2 \(\frac{p_\pi}{r^2}\)' .
\ea
Their time evolution does not provide any more constraints on the dynamics of $\mathcal{L}_2^{(0)}$. Overall, there are four constraints and the structure of the Poisson brackets between them reads
\be
	\{\phi_i(r), \phi_j(r')\} =
	\begin{pmatrix} 
		\; 0 & 0 & 0 & 0\\
		\; 0 & 0 & 0 & d(r, r')\\
		\; 0 & 0 & 0 & -d(r, r')\\
		\; 0 & -d(r, r') & d(r, r') & 0
	\end{pmatrix} ,
\ee
where $d$ is a nonzero distribution. Constraint $\phi_1$ is clearly a first-class constraint, while the matrix shows that a linear combination $\phi_2 + \phi_3$ is also a first-class constraint. The remaining two independent constraints $\phi_3, \phi_4$ are then second-class, which means the constraints in total remove three physical degrees of freedom. There were only three degrees of freedom in our problem described by $\mathcal{L}^{(0)}_2$, which means none of them is a physical degree of freedom. 
 This can also be seen directly from the  $\mathcal{L}^{(0)}_2$ Lagrangian itself.   Variation 
with respect to $A_r$ and $A_t$ produce constraints on $\pi$ rather than an equation of motion, in particular
$\dot\pi'=0$ or $\phi_4=0$.

In the special case of no vector background $v=0$, the decomposition of  Eq.~(\ref{NaiveSV}) in fact leads
to the same $\mathcal{L}_2^{(0)}$ as Eq.~(\ref{Lagr2Koyama}) since the additional term
from $\scalar$ enters as a total time derivative to ${\cal O}(m^0)$.  Thus, since $C=1$
\ba
	\mathcal{L}_2^{(0)} &=& \frac{3}{8} r^2 \sin \theta\(\pi'^2 - \dot \pi^2\),
\ea
which appears to be a normal kinetic term for $\pi$ that is a ghost in this $\alpha_3=\alpha_4=0$
theory and potentially healthy in other cases studied by Ref.~\cite{Koyama:2011yg}. 
 The vector component from $(A_t,A_r)$ appears to be a strongly coupled degree
of freedom with no kinetic term or coupling to the scalar at quadratic level.   
Although we are left with the correct answer of one degree of freedom, it does not
have the same dynamics  as the correct expansion in $m$ since the equation of motion for $\pi$ admits wavelike solutions.  Furthermore, the number of degrees of freedom would 
appear to be discontinuous as $v\rightarrow 0$, unlike in the correct analysis.

Thus for no $v$ does the Minkowski scaling limit provide the correct answer.  
The scaling with $m$ implied by canonical normalization of the degrees of freedom there does
not carry over to the self-accelerating solution where the kinetic structure begins at
first order in the curvature correction to the Minkowski limit.

\subsection{Gauge Fixing}
\label{sec:gauge}

 Ref.~\cite{Koyama:2011wx} in fact came to the conclusion that in the Minkowski scaling limit, the
 $v\ne 0$ case with a vector background propagates two degrees of freedom rather than
 the (also erroneous) zero degrees of freedom shown in the previous section.   
 We shall now show that that conclusion arises from fixing a gauge condition directly
 in the Lagrangian rather than at the level of the equations of motion.
 
  In the
 Minkowski scaling limit the introduction of the additional \stucky\ field $\pi$ 
 in Eq.~(\ref{NaiveSV}) restores  $U(1)$ gauge symmetry to the vector $A_\mu$.   
In this limit, we can take advantage of the gauge freedom to eliminate the non-dynamical
 field $A_t$.    In particular Ref.~\cite{Koyama:2011wx} chose a Lorenz gauge condition
 to demand that $A_\mu$ be divergence-free
 \be
	\label{DivergenceFree}
	\dot A_t - \frac{1}{r^2} \(r^2 A_r\)' = 0 .
\ee
 (We again stress that $A_t, A_r$ as defined are not components of a spacetime vector beyond the Minkowski limit.)
This condition can always be satisfied, because the Lagrangian \reff{Lagr2Koyama} is invariant under a $U(1)$ symmetry
 \be
	A_\mu \rightarrow A_\mu + \partial_\mu \varphi,
	\label{U1}
\ee
where $\varphi$ is an arbitrary scalar function.

If we use the divergence-free condition at the Lagrangian level, we can rewrite one of the terms as
\be
	\label{KoyamaFixing}
	\int dt\, dr\, r^2 \pi' \dot A_r \rightarrow \int dt\, dr\, r^2 \dot \pi \dot A_t ,
\ee
where we omitted unimportant numerical factors. Together with the term $\sim \dot \pi^2$ we then have two kinetic terms, which can be diagonalized to give two propagating degrees of freedom. This is in contradiction with the result of the previous section.

The operation which upset the counting of degrees of freedom is the use of the divergence-free condition \reff{KoyamaFixing} at the Lagrangian level.  This is not allowed, because the condition \reff{DivergenceFree} is a mere fixing of the gauge redundancy  brought about by the introduction of $\pi$ in Eq.~(\ref{NaiveSV}).

Given the $U(1)$ gauge symmetry, it is perhaps useful to illustrate the problem in the more
familiar setting of classical electromagnetism.   The Maxwell
Lagrangian
\be
	\label{LagrEM}
	\mathcal{L}^{(\mathrm{EM})} = - \frac{1}{4}\sqrt{-g} F_{\mu\nu}F^{\mu\nu}
\ee
with $F_{\mu\nu} = \partial_\mu A_\nu-\partial_\nu A_\mu$ propagates 
 two degrees of freedom due to presence of two first-class constraints. 
 Since it possesses the same $U(1)$ symmetry of Eq.~(\ref{U1}),
we can choose 
\be
	A_t = 0,
\ee
which defines the so-called temporal gauge.  
If we impose this condition on the Lagrangian level and drop all terms with $A_t$ in
the Minkowski limit of  Eq.~\reff{LagrEM}, we lose the constraint that it imposes.   The result is  a Lagrangian with three degrees of freedom rather than the correct two.
This is the same problem that occurs by gauge fixing the Minkowski scaling limit 
Lagrangian \reff{Lagr2Koyama} except  that the spherical symmetry assumption eliminates
the two correct degrees of freedom.

\section{Covariant  Perturbations}
\label{sec:vectors}

In \S \ref{sec:covariantform}, we construct a manifestly covariant form for the quadratic
Lagrangian for all \stucky\ and metric perturbations and all parameters of the massive gravity model extending the techniques of 
Ref.~\cite{Mirbabayi:2011aa}.   This form involves tensors constructed from the
self-accelerating background solution, obtained  in exact form for any
isotropic  solution, including inhomogeneous ones in \S \ref{sec:tensorform}.  We use these relations
to study the kinetic structure of the quadratic Lagrangian in the exact background and the locally
flat expansion in \S \ref{sec:kinetic}.   Finally in \S \ref{sec:koyamaform} we apply these general
results to the specific case studied in \S \ref{sec:decoupling}.

\subsection{Covariant Quadratic Lagrangian}
\label{sec:covariantform}

Given that the kinetic terms of the \stucky\ fields only appear at ${\cal O}(m^2)$ 
or equivalently as a curvature correction to the Minkowski limit,  $\phi^a$ cannot
be viewed as a spacetime vector
even for fluctuations around a locally flat patch.  Instead they transform as a vector
in the fiducial or internal space and as 4 scalars in the spacetime.    
Nonetheless by aligning the tetrad of the spacetime metric with the internal space by
a choice of vierbein,
we can construct objects from the \stucky\ scalars that transform as vectors
in the background spacetime  \cite{Mirbabayi:2011aa}.  From these objects we
can construct a manifestly covariant quadratic Lagrangian for the \stucky\  and metric perturbations.

In our case where we know the solution to the \stucky\ fields in the background, this
construction is particularly simple  \cite{Nibbelink:2006sz,Deffayet:2012zc,Gratia:2013gka}.   Given that
\begin{equation}
\ul{\bar\gamma}{\mu}{\alpha} \ul{\bar\gamma}{\alpha}{\nu} = \bar g^{\mu\alpha}\bar \fid_{\alpha\nu},
\end{equation}
and that $\partial_\mu \phi^a$ is an inverse vierbein of the fiducial metric $\Sigma_{\mu\nu}$,
it is easy to show that the quantity   $e^\mu_{\;\,a}$ constructed as the matrix manipulation of
\begin{equation}
\label{gammaDef}
\bar \gamma^{\mu}_{\;\,\nu} = e^\mu_{\;\,a} \partial_\nu\bar \phi^a
\end{equation}
is a vierbein of the background spacetime metric
\begin{equation}
\bar g_{\mu\nu}  e^\mu_{\;\,a}e^\nu_{\;\,b} = \eta_{ab}.
\end{equation}
Thus 
\begin{eqnarray}
\label{SigmaExpansion}
\Sigma^{\mu}_{\;\,\nu}  &=& g^{\mu\alpha}(e^\rho_{\;\,a}\partial_\alpha \phi^a)
(e^\sigma_{\;\,b}\partial_\nu \phi^b)\bar g_{\rho\sigma}
\end{eqnarray}
is constructed out of an object that now transforms as a tensor in the background spacetime \begin{eqnarray}
\label{GammaExpansion}
\bar g_{\mu\sigma}  e^\sigma_{\;\,a} \partial_\nu \phi^a& = &
\bar g_{\mu\sigma}  e^\sigma_{\;\,a} \partial_\nu (\bar \phi^a + \delta\phi^a) \nonumber\\
&=& \bar \gamma_{\mu \nu} + \bar g_{\mu\sigma}  e^\sigma_{\;\,a} \partial_\nu  (\delta\phi^a) .
\end{eqnarray}
Note that we raise and lower spacetime indices with the background metric to leading order. 
Although this quantity transforms as a spacetime tensor, its relation to the spacetime vector built out of the \stucky\ fields
\begin{equation}
\aM^\mu =  e^\mu_{\;\,a}\delta \phi^a
\end{equation}
requires the introduction of  connection coefficients \cite{Mirbabayi:2011aa}
\begin{eqnarray}
\label{IntroduceStuckPert}
\bar g_{\nu\sigma}  e^\sigma_{\;\,a} \partial_\mu  (\delta\phi^a)&=&
\partial_\mu \aM_\nu 
- C^{\sigma}_{\mu\nu} \aM_\sigma \nonumber\\
&=& \aM_{\nu;\mu} -  [C^{\sigma}_{\mu\nu}- \Gamma^{\sigma}_{\mu\nu}] \aM_\sigma ,
\end{eqnarray}

Here $\Gamma$ is the usual Christoffel symbol  formed from $\bar g_{\mu\nu}$ and
defines covariant derivatives or parallel transport of vectors in the spacetime.   
$C$ is the connection coefficient associated with the space-time dependence of
the alignment of the internal space and tetrad encapsulated by the change in the vierbein
\begin{equation}
C^{\sigma}_{\mu\nu} = \partial_\mu (\bar g_{\nu\lambda} e^\lambda_{\;\,a}) \bar g^{\rho\sigma} 
[e^{-1}]^{a}_{\;\,\rho} .
\end{equation}
Note that
\be
	C'^\sigma_{\mu\nu} = C^{\sigma}_{\mu\nu}-\Gamma^{\sigma}_{\mu\nu}
\ee
is the difference of two connections and thus transforms as a tensor even though connection coefficients do not. 

We can now characterize the quadratic Lagrangian of the gravitational sector 
in terms of these variables.  For notational simplicity we divide the Lagrangian into
terms involving the \stucky\ fields,  
metric perturbations
\be
	g_{\mu\nu} = \bar g_{\mu\nu} + h_{\mu\nu},
\ee
and cross terms.   For convenience we factor out common terms following the spherically
symmetric
results of Ref.~\cite{Wyman:2012iw} and break the terms into component pieces
\begin{eqnarray}
{\cal L}_{2} &=& P_1'(x_0) m^2 \Mpl^2 \sqrt{-\bar g}\left( { L}_\mathit{SS} + { L}_{Sh} + { L}_{hh}\right) \nonumber\\
&&+ {\cal L}_{hh}^{\rm (EH)} + \mathcal{L}^{(\Lambda)}_{hh},
	\label{allquad}
\end{eqnarray}
where $P_1'(x_0)$ is a model parameter dependent constant whose definition we
will give in Eq.~(\ref{eqn:P1}).  

The quadratic Lagrangian for the pure \stucky\ terms 
should then take the form 
\begin{equation}
\label{MirbabayiForm}
L_{SS}
= B^{\mu\nu\alpha\beta} [  \aM_{\nu;\mu} - C'^{\rho}_{\mu\nu}
\aM_{\rho}  ][\aM_{\beta;\alpha} - C'^{\sigma}_{\alpha\beta}
\aM_{\sigma} ],
\end{equation}
where $B$ is a tensor formed from background quantities $\bar g_{\mu\nu}, \bar\gamma_{\mu\nu}$. Unlike Ref.~\cite{Mirbabayi:2011aa} we factor out $\sqrt{-\bar g}$ so that $B$
transforms as a tensor.

We can similarly determine the functional form of the coupling of
the \stucky\ fields to the metric perturbation
\be
	L_{Sh} = D^{\mu\nu\alpha\beta }h_{\mu\nu} \(\aM_{\beta;\alpha} - {C'}_{\alpha\beta}^\sigma \aM_\sigma\),
	\label{stuckymetric}
\ee
where $D$ is a tensor constructed out of the background quantities.

Finally, the Lagrangian quadratic in the metric perturbations can be split into a part coming from the Einstein-Hilbert action and a part coming from  massive gravity.
The Einstein-Hilbert piece takes the same form as in general relativity (e.g.~\cite{Natsuume:2010ky,Gullu:2010em})
\begin{eqnarray}
\frac{ \mathcal{L}_{hh}^{\rm  (EH)} }{\sqrt{-\bar g} \Mpl^2 }&=&\left( \frac{1}{2} h^{\mu\alpha}\lu{h}{\alpha}{\nu} - \frac{1}{4}  h h^{\mu\nu} \right) \bar R_{\mu\nu}  \\
&& +
 \left(\frac{1}{16}  h^2 - \frac{1}{8} h_{\mu\nu} h^{\mu\nu} \right) \bar R -\frac{1}{8} h^{\mu\nu;\alpha} h_{\mu\nu;\alpha} \nonumber\\
&&
 + \frac{1}{4} h^{\mu\nu;\alpha} h_{\nu\alpha;\mu}
+\frac{1}{8} h_{;\alpha} h^{;\alpha} 
-\frac{1}{4} \ul{h}{\mu\nu}{;\nu} h_{;\mu} ,\nonumber
\end{eqnarray}
where $h= \ul{h}{\alpha}{\alpha}$. 
For the massive gravity metric-metric terms, first we have the  term that depends on the effective cosmological constant of
the self-accelerating background which represents a non-dynamical 
change in the measure.  To see this note that a true cosmological constant has
a  contribution to the action of
\be
{\cal L}^{(\Lambda)} =- \Mpl^2 \sqrt{-g} \Lambda ,
\ee
and its non-dynamical quadratic metric terms are given by the expansion
\be 
\sqrt{-g} \approx \sqrt{-\bar g} \left[ 1 + \frac{1}{2} h + \frac{1}{2} \left( \frac{1}{4} h^2 - \frac{1}{2} h^{\mu\nu}h_{\mu\nu} \right)\right]
\ee 
as
\be
\frac{\mathcal{L}^{(\Lambda)}_{hh}}{\sqrt{-\bar g}\Mpl^2} =  \left(\frac{1}{4} h_{\mu\nu} h^{\mu\nu}  -\frac{1}{8}  h^2  \right)\Lambda .
\ee
This piece will cancel terms in the Einstein-Hilbert Lagrangian by virtue of the
Einstein equations in the background.   We shall see this feature explicitly in the
construction of the perturbed stress energy tensor below.
The remaining massive gravity terms can be parameterized as
%
\ba
	L_{hh} &=&
 E^{\mu\nu\alpha\beta }h_{\mu\nu} h_{\alpha\beta},
	\label{metricmetric}
\ea
where  $E^{\mu\nu\alpha\beta}$ is a tensor that depends on the background quantities.

This completes the general description of the structural form for the covariant
quadratic Lagrangian derived of the gravitational sector.
 We now turn to the construction of the background tensors
$B$, $D$, $E$.

\subsection{Fluctuations around Isotropic Backgrounds}
\label{sec:tensorform}

For all  isotropic background solutions on the self-accelerating branch \cite{Gratia:2012wt}, 
there is a single universal form for the relationship between $B$, $D$, $E$ and the
 background tensors $\bar g_{\mu\nu}$, $\bar \gamma_{\mu\nu}$.
This includes the vacuum self-accelerating solutions of Ref.~\cite{Koyama:2011yg}
as well as its approximation in conformal coordinates, Eq.~(\ref{conformal}) that was
considered in \S \ref{sec:decoupling}.  It also includes the special cases of the
open self-accelerating solution \cite{Gumrukcuoglu:2011ew} which is known to
propagate no extra degrees of freedom from the mass term at quadratic level.

We therefore utilize the general construction of Ref.~\cite{Gratia:2012wt}.
As some aspects of this construction will be useful for extracting the background tensors, we review its
salient features  here. 
Any spherically symmetric metric can be written  in isotropic coordinates as
\ba
	\bar{g}_{\mu\nu}\dd x^\mu \dd x^\nu &=& -b^2(t,r) \dd t^2 \
	+ a^2(t,r) \big(\dd r^2 + r^2 \dd \theta^2 	\nonumber\\
	& & 
	+ r^2 \sin^2\theta \dd \phi^2\big),
\ea
whereas  the background \stucky\ fields can again be given by the isotropic form of Eq.~(\ref{stuckyback}).   The spacetime metric is diagonal in these coordinates and $\bar g^{\mu\alpha}\bar \Sigma_{\alpha\nu} = \ul{\bar\gamma}{\mu}{\alpha} \ul{\bar\gamma}{\alpha}{\nu}$ has off diagonal entries only
in the $(t,r)$ cases.    It is convenient to use matrix notation
here and so we define the $(t,r)$ block as 
\ba
	\bar\bgamma_2 & \equiv & 
	\begin{pmatrix}
		\ul{\bar\gamma}{t}{t}  &\ul{\bar\gamma}{t}{r} \\
		\ul{\bar\gamma}{r}{t}  & \ul{\bar\gamma}{r}{r}\\
	\end{pmatrix}.
\ea
Note that although $\gamma_{\mu\nu}$ is symmetric $\ul{\gamma}{\mu}{\nu}$ is not.  Its square
is related to the background \stucky\ fields as
\ba
			\bar\bgamma_2 \bar\bgamma_2  &=& 
	\begin{pmatrix}
		\dfrac{\dot T^2 -\dot R^2}{b^2}   &\dfrac{\dot T T' - \dot R R'}{b^2} \\
		\dfrac{\dot R R'  -\dot T T'}{a^2} & \dfrac{R'^2 - T'^2}{a^2}\\
	\end{pmatrix}. 
\ea
The general solution to the matrix square root is given by the Cayley-Hamilton theorem
\begin{equation}
[\bar\bgamma_2] \bar\bgamma_2 = \bar\bgamma_2\bar\bgamma_2 + ({\rm det}\bar\bgamma_2)
{\bf I}_2,
\label{CH}
\end{equation}
where ${\bf I}_2$ is the $2\times 2$ identity matrix.  The determinant can be written in terms of the determinant of the square of the matrix
 and hence in terms of the \stucky\ 
background
\begin{equation}
\label{DetGamma2}
{\rm det}\bar\bgamma_2 = \frac{ \dot T R' - \dot R T'}{ a b},
\end{equation}
and the trace similarly by taking the trace of Eq.~(\ref{CH}).
 Using this solution in the Lagrangian, we obtain the equations of motion for the background \stucky\ fields and find that on the self-accelerating branch 
\begin{eqnarray}
R(t,r) &=& x_0 r a(t,r), \nonumber\\
\, [ \bar\bgamma_2 ] &=& \frac{1}{x_0} {\rm det}\bar\bgamma_2 + x_0 ,
\label{SEOM}
\end{eqnarray}
where $x_0$ is a constant that solves $P_1(x_0)=0$ with
\begin{equation}
P_1(x) = 2 (3 -2 x)  +  6(x-1)(x-3)\alpha_3 +   24(x-1)^2 \alpha_4.
\label{eqn:P1}
\end{equation}
The second equation may be rewritten as an equation of motion for $T(t,r)$
\ba
\label{EOM}
	b^2 T'^2 + 2 a r(a' \dot T^2 - \dot a \dot T T')+r^2 (a' \dot T - \dot a T')^2 \nonumber
	\\
	= x_0^2 \(a'^2 b^2 r^2 + 2 a' a b^2 r - \dot a^2 a^2 r^2\) .
\ea
These solutions then require 
\begin{equation}
\ul{\bar\gamma}{\theta}{\theta}= \ul{\bar\gamma}{\phi}{\phi} = x_0.
\end{equation}
Note that in terms of the massive gravity parameter dependence  $T, R, {\bar\bgamma}\propto x_0$.

These solutions imply an effective cosmological constant in the background stress energy tensor
\begin{equation}
\bar T_{\mu\nu} =-\Lambda \Mpl^2 \bar g_{\mu\nu}= -\frac{1}{2}P_0(x_0) m^2 \Mpl^2 \bar g_{\mu\nu},
\label{bart}
\end{equation}
where
\begin{align}
P_0(x) &= - 12 - 2 x(x-6) - 12(x-1)(x-2)\alpha_3 
\nonumber\\&\qquad -24(x-1)^2\alpha_4.
\end{align}
Knowing $\bar \bgamma$ and the background \stucky\ fields, we can construct the vierbein $\ul{e}{\mu}{a}$ by solving Eq.~(\ref{gammaDef}).

 One useful consequence of
Eqs.~(\ref{CH}) and (\ref{SEOM}) is that a certain combination of $\bar g_{\mu\nu}$ and
$\bar \gamma_{\mu\nu}$ 
\be
\label{OurG}
	\bar\OurG_{\mu\nu} = \frac{1}{x_0} \bar \gamma_{\mu\nu} - \bar g_{\mu\nu}
\ee
obeys special properties.
First note that it is independent of the model parameter choices for $m,\alpha_3,\alpha_4$.  It is only non-zero for  $\bar \OurG_{tt}, \bar \OurG_{tr}$ and $\bar \OurG_{rr}$.
  Defining again this $2\times2$ block with upper and lower indices
as ${\bOurG}_2$ these components satisfy
\begin{equation}
[\bar\bOurG_2]\bar \bOurG_2 = \bar\bOurG_2 \bar \bOurG_2,
 \label{gammarelation}
\end{equation}
or equivalently ${\rm det}(\bar \bOurG_2)=0$.  More explicitly, after lowering
indices
\be
	\bar\OurG_{tt} \bar\OurG_{rr} = \bar\OurG_{tr}^2 .
	\label{offdiagonalrelation}
\ee
Likewise we can write Eq.~(\ref{gammarelation}) in $4\times 4$ index notation as
\ba
\label{gammarelationcomponents}
	\bar\OurG_{\mu\nu} \bar g^{\nu\alpha} \bar \OurG_{\alpha\beta} = [\bar\bOurG] \bar \OurG_{\mu \beta} .
\ea

We can now construct the tensors $B$, $E$ and $D$ from these background tensors, 
specifically $\bar g_{\mu\nu}$ and $\bar \OurG_{\mu\nu}$.
Beginning with $B$, 
we need to determine the perturbation to the $\ul{\gamma}{\mu}{\nu}$ solution given a 
\stucky\ perturbation in its square $\ul{\Sigma}{\mu}{\nu}$.  
To determine values of components $B^{\mu\nu\alpha\beta}$ it is sufficient to keep track of  the coefficient of $\aM_{\nu;\mu}\aM_{\beta;\alpha}$ in the expansion of the Lagrangian Eq.~(\ref{drgt}) to second order in the perturbations $\aM$. This in turns means we need expansion of $\ul{\gamma}{\mu}{\nu}$ to second order in $\aM$.

We start with the defining relation
\be
	\ul{\Sigma}{\mu}{\nu} = \ul{\gamma}{\mu}{\alpha} \ul{\gamma}{\alpha}{\nu}
\ee
and expand the tensor $\Sigma$ order by order in $\aM$,
\be
	\ul{\Sigma}{\mu}{\nu} = \ul{\bar \Sigma}{\mu}{\nu} + \ul{{\Sigma^{(1)}}}{\mu}{\nu} + \ul{{\Sigma^{(2)}}}{\mu}{\nu} + \dots,
\ee
and similarly for $\gamma$. The various $\ul{{\Sigma^{(i)}}}{\mu}{\nu}$ can be directly obtained in terms of background quantities $\bargamma, \bar g$ and $\aM$ from Eqs.~(\ref{SigmaExpansion}), (\ref{GammaExpansion}) and (\ref{IntroduceStuckPert}).

Using the zeroth order solution 
\be
	\ul{\bar \Sigma}{\mu}{\nu} =\ul{\bargamma}{\mu}{\alpha} \ul{\bargamma}{\alpha}{\nu},
\ee
we match orders as
\ba
	\ul{{\Sigma^{(1)}}}{\mu}{\nu}	&=& \ul{\bargamma}{\mu}{\alpha}\ul{{\gamma^{(1)}}}{\alpha}{\nu} + \ul{{\gamma^{(1)}}}{\mu}{\alpha}\ul{\bargamma}{\alpha}{\nu}, \\
	\ul{{\Sigma^{(2)}}}{\mu}{\nu}	&=& \ul{\bargamma}{\mu}{\alpha}\ul{{\gamma^{(2)}}}{\alpha}{\nu}+ \ul{{\gamma^{(1)}}}{\mu}{\alpha}\ul{{\gamma^{(1)}}}{\alpha}{\nu} + \ul{{\gamma^{(2)}}}{\mu}{\alpha}\ul{\bargamma}{\alpha}{\nu} .
	\nonumber
\ea
Each order represents 16 linear equations for components of $\gamma^{(i)}$ and can be readily solved iteratively.


With the explicit form for $\gamma$ up to second order in \stucky\ perturbations, we can perturb the Lagrangian density $\mathcal{L}^{\rm (MG)}$ and read off 
$L_{SS}$.  The coefficients of the various terms form the $B$-tensor.  
For a general spherically symmetric background solution, it is possible to express these components in terms of the background tensors $\bar \OurG_{\mu\nu}$ and $\bar g_{\mu\nu}$. Because of the relation (\ref{gammarelationcomponents}) and definition (\ref{OurG}), all tensor structures involving more than two gamma matrices contracted with an inverse metric such as
\be
	\bargamma_{\alpha\beta}\bar g^{\beta \kappa} \bargamma_{\kappa \rho} \bar g^{\rho \sigma} \bargamma_{\sigma \delta}
\ee
can be written as a linear combination of $\bar\OurG_{\alpha\delta}$, $ \bar g_{\alpha\delta}$ with coefficients which are spacetime scalars built out of traces $[\bar\bOurG]$. This relation greatly reduces the number of terms we have to take into account as only 12 of them are in principle independent, such as $\bar g_{\mu\nu}\bar g_{\alpha\beta}$ and $\bar \OurG_{\mu \beta} \bar g_{\nu \alpha}$. Coefficients in front of these terms must be spacetime scalars, which must be functions of the trace $[\bar\bOurG]$. In principle these scalars would depend also on $\ul{\bargamma}{\alpha}{\beta}\ul{\bargamma}{\beta}{\alpha}, \ul{\bargamma}{\alpha}{\beta}\ul{\bargamma}{\beta}{\rho}\ul{\bargamma}{\rho}{\alpha} , \dots$ but in the present case these can be expressed as functions of $[\bar\bOurG]$ by a (repeated) use of Eq.~(\ref{gammarelationcomponents}). Taking into account the relation  (\ref{offdiagonalrelation}) for the off-diagonal elements $\bar \OurG_{tr}$, it is possible to reduce the coefficients in front of the tensorial structures into simple forms.

The \stucky-\stucky\ Lagrangian obtained in this manner can be written as
\begin{eqnarray}
L_{SS} = ( \tilde B^{\mu\nu\alpha\beta} - \Delta B^{\mu\nu\alpha\beta} )
 [  \aM_{\nu;\mu} - C'^{\rho}_{\mu\nu}
\aM_{\rho}  ][\aM_{\beta;\alpha} - C'^{\sigma}_{\alpha\beta}].
\nonumber\\
\end{eqnarray}
Here we have separated out a term that is a total derivative and hence may be
dropped from the Lagrangian
\ba
	\Delta B^{\mu\nu\alpha\beta} &=& \frac{x_0 P_2'(x_0)}{8 P_1'(x_0)} \Big[ (1 + [\bar \bOurG]) (g^{\mu\nu}g^{\alpha\beta} - g^{\mu\beta}g^{\alpha\nu}) \\
	&&
	+ (\chi^{\mu\beta}g^{\nu\alpha} + \chi^{\nu\alpha}g^{\mu\beta} - \chi^{\mu\nu}g^{\beta\alpha} - \chi^{\alpha\beta}g^{\nu\mu})\Big] \nonumber
\ea
with
\be
P_2(x) = -2+ 12 \alpha_3(x - 1) - 24 \alpha_4(x - 1)^2
\ee
from the dynamical piece which itself can be broken up into terms that are symmetric and antisymmetric
in permutation of indices
\begin{equation}
\tilde B^{\mu\nu\alpha\beta} = \tilde B^{(\mu\nu)(\alpha\beta)}  + \tilde B^{[\mu\nu][\alpha\beta]}  ,
\end{equation}
where	
\ba
\label{BExplicit}
	\tilde B^{(\mu\nu)(\alpha\beta)} &=& -\frac{1}{8} \( \bar g^{\mu\nu} \bar g^{\alpha \beta} - \frac{1}{2} \bar g^{\mu \alpha} \bar g^{\nu \beta} - \frac{1}{2} \bar g^{\mu \beta} \bar g^{\nu \alpha}\),\nonumber\\
	\tilde B^{[\mu\nu][\alpha\beta]} &=&   \frac{1}{16} \(\bar \OurG^{\mu \alpha}\bar g^{\nu \beta} + \bar \OurG^{\nu \beta}\bar g^{\mu \alpha} - \bar \OurG^{\mu \beta}\bar g^{\nu \alpha} - \bar \OurG^{\nu \alpha}\bar g^{\mu \beta}\)	\nonumber\\
	&& -
	\frac{1+ [\bar \bOurG]}{16}\(\bar g^{\mu\alpha}\bar g^{\nu\beta} - \bar g^{\mu\beta}\bar g^{\nu\alpha}\).
\ea	

We can form an alternate 
 representation of the tensor $B$ by removing any combination of the total derivative term.
 In particular the form
 \begin{equation}
 B^{\mu\nu\alpha\beta} = \tilde  B^{\mu\nu\alpha\beta} + \frac{P_1'(x_0)}{x_0 P_2'(x_0)}  \Delta B^{\mu\nu\alpha\beta} ,
 \end{equation}
 or explicitly
 \ba
	B^{\mu\nu\alpha\beta}&=&
		\frac{ [\bar\bOurG]}{8}\( \bar g^{\mu\nu} \bar g^{\alpha \beta}-\frac{1}{2} \bar g^{\mu\beta}\bar g^{\nu\alpha} - \frac{1}{2} \bar g^{\mu \alpha} \bar g^{\nu \beta} \)
	\nonumber\\
	&&+ \frac{1}{16} \(\bar g^{\mu\alpha} \bar\OurG^{\nu \beta} + \bar g^{\nu\beta} \bar\OurG^{\mu\alpha} + \bar g^{\mu \beta} \bar\OurG^{\nu\alpha} + \bar g^{\nu \alpha} \bar\OurG^{\mu\beta}\)
	\nonumber\\
	&&- \frac{1}{8}\(\bar g^{\mu\nu}\bar\OurG^{\alpha\beta} + \bar g^{\alpha\beta} \bar\OurG^{\mu\nu}\),
	\label{eqn:Bnew}
\ea
is useful as we shall see below.
To keep these representations distinct we reserve the $B$ tensor symbol for this form.
Note that it is symmetric under the exchange of the  first or last two indices.

Similarly  we can determine expressions for $D^{\mu\nu\alpha\beta}$ for the
\stucky-metric terms of Eq.~(\ref{stuckymetric}) and $E^{\mu\nu\alpha\beta}$ for the metric-metric terms of Eq.~(\ref{metricmetric})
\ba
	D^{\mu\nu\alpha\beta}&=& - x_0 B^{\mu\nu\alpha\beta},
	\nonumber\\
	E^{\mu\nu\alpha\beta} &=&  \frac{x_0^2}{4} B^{\mu\nu\alpha\beta}.
	\label{DEexplicit}
\ea
Note that these expressions contain contributions from varying both $\sqrt{-g}$ and $\Sigma$ with respect to the metric. 

Thus the whole quadratic Lagrangian can be written very compactly as
\ba
\label{SimplifiedLagr}
	\mathcal{L}_2  =\mathcal{L}_{hh}^{\rm (EH)}+\mathcal{L}_{hh}^{(\Lambda)} + P_1'(x_0)m^2 \Mpl^2 \sqrt{-\bar g} B^{\mu\nu\alpha\beta} W_{\mu\nu} W_{\alpha\beta},\nonumber\\
\ea
where
\begin{equation}
W_{\mu\nu} = \aM_{\nu;\mu} - C'^{\rho}_{\mu\nu}  \aM_\rho - \frac{x_0 h_{\mu\nu}}{2} .
\end{equation}
This result represents the full quadratic Lagrangian of the gravitational sector
 in any 
isotropic self-accelerating branch solution of the theory.
It trivially allows the addition of minimally coupled matter but does not necessarily hold beyond the isotropic
assumption.  Interestingly this includes the case where the background spacetime metric
is exactly Minkowski due to the canceling impact of a bare cosmological constant.   This case is still
not the same as the Minkowski decoupling limit, since self-accelerating branch solutions 
always have non-trivial \stucky\ backgrounds given by Eq.~(\ref{SEOM}).  As we shall see, 
this generalizes the result of the previous section, that the locally flat expansion of
a self-accelerating solution is not the same as the  Minkowski decoupling limit.

With the explicit formulae for $D$ and $E$ and expansions (\ref{stuckymetric}), (\ref{metricmetric}) we can also construct the linear fluctuations in the stress energy tensor away from the self-accelerating background of Eq.~(\ref{bart}),
\ba
	\delta \lu{T}{\mu}{\nu} &=& - \frac{2}{\sqrt{- g}}g^{\nu\alpha} \frac{\delta \mathcal{L}_{\rm MG}}{\delta g^{\mu\alpha}} - \lu{\bar T}{\mu}{\nu}  \\
	&\approx& -2 P_1'(x_0) m^2 \Mpl^2 x_0 \lu{B}{\mu}{\nu\alpha\beta}
	W_{\alpha\beta}.\nonumber
	 \ea

	 {It is now clear why we grouped terms in Eq.~(\ref{metricmetric}).  Since the
	 stress energy fluctuation is the source of $h_{\mu\nu}$ through the Einstein equations, these are the only terms with  dynamical impact on the metric.}
The stress tensor constructed in this way through $D$ and $E$ agrees with the expansion of the exact result \cite{Gratia:2012wt} and serves as a check on their derivation. 
Note that the equations of motion derived from the quadratic Lagrangian satisfy 
covariant conservation of the massive gravity stress-energy tensor $\nabla^\mu T_{\mu\nu}=0$ regardless of the matter content.

\subsection{Kinetic Structure}
\label{sec:kinetic}

Although the quadratic  Lagrangian of Eq.~(\ref{SimplifiedLagr})  with the explicit form for 
 the background $B$ tensor of Eq.~(\ref{eqn:Bnew}) is
 complete, its implication for the dynamics of the \stucky\ fields $\aM^\mu$ is not
 yet explicit.  
 Is is therefore useful to further isolate
the pieces associated with Maxwell type terms involving the antisymmetric field strength tensor
\begin{equation}
f_{\mu\nu} = \aM_{\nu;\mu} - \aM_{\mu;\nu}
\end{equation}
and reorganize  the terms in ${L}_{SS}$ by the number of appearances of the
field strength tensor
\begin{equation}
 L_{SS} =   { L}_\mathit{ff} + { L}_{f\aM} + { L}_{\aM\aM}.
\end{equation}

Reducing the Lagrangian to this form is simpler in the $\tilde B$ representation of
Eq.~(\ref{BExplicit}).  
First note that we can add total derivatives to rewrite
\ba
	&& \tilde B^{(\mu\nu)(\alpha\beta)} \aM_{\nu;\mu}\aM_{\beta;\alpha}  + \frac{1}{8} \(\aM^\mu \ul{\aM}{\nu}{;\nu}\)_{;\mu} - \frac{1}{8} \(\aM^\mu \ul{\aM}{\nu}{;\mu}\)_{;\nu}  \nonumber\\
	\nonumber\\
	&&\quad= \frac{1}{16} \aM_{\mu;\nu} \aM^{\mu;\nu} - \frac{1}{16} \aM_{\mu;\nu} \aM^{\nu;\mu} 
	- \frac{1}{8} \aM^\mu \(\ul{\aM}{\nu}{;\mu\nu}-\ul{\aM}{\nu}{;\nu\mu}\)\nonumber\\
	&&\quad  =\frac{1}{32} f_{\mu\nu}f^{\mu\nu}- \frac{1}{8} \aM^\mu \aM^\sigma \bar R_{\sigma\mu} .
\ea
$\bar R_{\sigma\mu}$ denotes the usual Ricci tensor built out of the background metric $\bar g$.


After similar integrations by parts, we arrive at the result
\begin{eqnarray}
\label{GeneralffTerm}
	L_\mathit{ff} &=& - \frac{1}{32} [\bar \bOurG] f_{\mu\nu} f^{\mu\nu} + \frac{1}{16}\ul{\bar\chi}{\nu}{\beta} f_{\mu\nu} f^{\mu\beta}, \nonumber\\
		L_{f\aM} &=& {C'}_{\mu\nu}^\alpha  (2\tilde B^{(\mu\nu)(\rho\sigma)} \aM_\sigma f_{\rho\alpha }   - \tilde B^{[\mu\nu][\rho\sigma]} \aM_\alpha f_{ \rho\sigma}) ,\nonumber\\
	L_{\aM\aM} &=& 
	\Big[ 
	 2 \tilde B^{(\mu\nu)(\alpha\sigma)}C'^\rho_{\mu\nu;\alpha} 
	 - \tilde B^{(\mu\nu)(\rho\sigma)}C'^\alpha_{\mu\nu;\alpha} \nonumber\\
	 && + \tilde B^{\mu\nu\alpha\beta} C'^\rho_{\mu\nu}  C'^\sigma_{\alpha\beta}  -\frac{1}{8} \bar R^{\rho\sigma} \Big]
	 \aM_\sigma \aM_\rho.
	\end{eqnarray}
	
In simplifying the expressions we have integrated by parts and used the fact that $\tilde B^{(\mu\nu)(\alpha\beta)}_{\phantom{(\mu\nu)(\sigma\rho)};\rho = 0}$ as it is constructed from products of the metric in Eq.~(\ref{BExplicit}). 

The only place that time derivatives appear in the $SS$ terms are in the $\mathit{ff}$ and $f\aM$ pieces.
Given the antisymmetry of $f_{\mu\nu}$ it is clear that the field $\aM_t$ is nondynamical
reflecting  the
absence of the Boulware-Deser ghost.
This  structure of the \stucky\ Lagrangian is expected based on general theoretical arguments \cite{Mirbabayi:2011aa}.

Now consider the $\mathit{ff}$ terms that would usually provide quadratic kinetic terms and hence
second order equations of motion.
Inspection of Eq.~(\ref{GeneralffTerm}) shows
that the $f_{tr}^2$ term always vanishes identically.
This in turn means that around spherically symmetric solutions, there is no Maxwell term for spherically symmetric perturbations, which is in full agreement with the investigations of previous sections and with the full theory \cite{Wyman:2012iw,Khosravi:2013axa}. 

The terms $f_{t\theta}^2, f_{t\phi}^2$ have coefficients that are proportional to
\be
		 \bar\OurG_{rr} \propto \frac{ R'^2 - T'^2}{x_0^2} - a^2 ,
		 \label{fttheta}
\ee
and give Maxwell-like kinetic terms to the transverse modes when non-vanishing.
Note that for the special open universe solution of Ref.~\cite{Gumrukcuoglu:2011ew}, the fiducial metric is diagonal in isotropic coordinates and this quantity vanishes.   Thus the 
strongly-coupled anisotropic modes of that model is an artifact of this special symmetry 
that is imposed.


The remaining kinetic terms are first order.   From the $\mathit{ff}$ term, we have the mixed terms $f_{t\theta}f_{r\theta}$, $f_{t\phi}f_{r\phi}$ which appear with coefficients proportional to
\be
\label{ftfrRelation}
	\bar \OurG_{tr} = \sqrt{\bar \OurG_{rr} \bar \OurG_{tt}}.
\ee
Compared with Eq.~(\ref{fttheta}), this means  that the mixed terms will
scale differently from the pure kinetic Maxwell terms due by a factor of $\sqrt{ \bar\OurG_{tt}/\bar \OurG_{rr}}$.

The $\mathit{f\aM}$ terms have a general structure
\ba
	L_{f\aM} &=& K_1 \(\aM_\theta f_{r\theta} \sin^2 \theta + \aM_\phi f_{r\phi}\)
	\nonumber \\ 
	&& + K_2 \(\aM_\theta f_{t\theta} \sin^2 \theta + \aM_\phi f_{t\phi}\)
	\nonumber \\
	&& + K_3 \aM_t f_{tr} + K_4 \aM_r f_{tr} .
\ea 
The last two coefficients can be rewritten in a succinct form
\ba
	K_3 &=& \frac{b R' {\rm det}\bar\bgamma_2 - x_0^2 a \dot T }{2 x_0 \tr{\bar\bgamma_2}a^2 b^3 R},\nonumber\\
	K_4 &=& \frac{x_0^2 b T' - a \dot R {\rm det}\bar\bgamma_2}{2x_0 \tr{\bar\bgamma_2}a^3 b^2 R},
\ea
while the expressions for $K_1, K_2$ are more involved and will not be given here.
Note that the $K_1$ term is nondynamical as is $K_2$ and $K_4$ since, e.g.
\ba
	\aM_\phi f_{t\phi} &=& \frac{1}{2}\frac {\partial \aM_\phi^2}{\partial t}   -\aM_\phi \aM_{t,\phi}
\ea
so that the time derivative can be moved onto the background by integration by parts.
For the special case of the open universe solution \cite{Gumrukcuoglu:2011ew}, $K_3=0$
and combined with the angular terms this means that all 3 \stucky\ fields are non-dynamical.

In the general case \stucky\ dynamics are supplied by the terms $f_{t\theta}^2$, $f_{t\phi}^2$, $f_{t\theta}f_{r\theta}$, $f_{t\phi}f_{r\phi}$ and $\aM_t f_{tr}$.
It is interesting to generalize the considerations of \S\ref{sec:conformalkinetic} for fluctuations
around a locally flat patch to see at what order in curvature corrections that each contributes.  
Any isotropic metric can be considered locally as Minkowski plus curvature corrections 
and hence 
\ba
	a, b &=& 1 + \mathcal{O}(m^2).
\ea
{For notational simplicity we have here assumed vacuum self-acceleration cases here;
more generally we would replace $\mathcal{O}(m^2)$ with $\mathcal{O}(\bar R)$.}
Thus given Eq.~(\ref{SEOM}) for the exact solution, we may approximate
\ba
	R &=& x_0 r + \mathcal{O}(m^2) .
\ea
The other \stucky\ equation of motion (\ref{EOM}) then implies
\be
	T'^2 = \mathcal{O}(m^2)    
\ee
which means the unitary gauge time $T$ does not depend on the spatial coordinate in the leading order, $T = T(t) + \mathcal{O}(m)$.
With this solution, we can write down the components of the background tensor~$\OurG$
\be
	\OurG_{\alpha \beta} = 
	\begin{cases}
		1 - \dot T/x_0 + \mathcal{O}(m) &\mbox{if } \alpha = t, \beta = t\\
		\mathcal{O}(m) &\mbox{if } \alpha = t, \beta = r\\
		\mathcal{O}(m^2) &\mbox{if } \alpha = r, \beta = r\\
		0 & \mbox{otherwise}
	\end{cases}.
	\label{bargammaexp}
\ee
 From Eq.~(\ref{fttheta}) it follows that
the kinetic Maxwell terms $f_{t\theta}^2$ and $f_{t\phi}^2$ are at most $\mathcal{O}(m^2)$.
The leading order
 kinetic $\mathit{ff}$ terms are $f_{t\theta}f_{r\theta}$, $f_{t\phi}f_{r\phi}$ which appear already at order $\mathcal{O}(m)$ due to the square root in Eq.~(\ref{ftfrRelation}) and $\mathcal{O}(m^2)$ suppression of~$\OurG_{rr}$. 

From Eq.~(\ref{DetGamma2})
\be
	{\rm det}\bar\bgamma_2 = x_0 \dot T + \mathcal{O}(m^2),
\ee
and so $K_3\aM_t f_{tr}$ also starts at  most at $\mathcal{O}(m^2)$.

On the other hand, the spatial derivative terms in the Lagrangian do not necessarily
begin at suppressed orders.   We find that the space-space Maxwell terms can have contributions at ${\cal O}(m^0)$
\ba
L_\mathit{ff} &=&
  \frac{1}{16}\(1 - \frac{\dot T}{x_0}\) \(f_{r\theta}f^{r\theta} + f_{r\phi}f^{r\phi} + f_{\phi \theta}f^{\phi\theta}\)\nonumber\\ && + \mathcal{O}(m).
\ea

For the case with $\dot T = x_0 + \mathcal{O}(m)$, the terms $2 a r a' \dot T^2$, $2 x_0^2 a' a b^2 r$ in the equation of motion cancel in the leading order and we are left with 
\be
	T'^2 = \mathcal{O}(m^4)  .   
\ee
This means that in fact $T = x_0 t + \mathcal{O}(m^2)$ and
\be
	\OurG_{\alpha \beta} = \mathcal{O}(m^2) .
	\label{bargammav0}
\ee
In this case, which corresponds to $v= 0$ in the example of \S \ref{sec:decoupling},
 all $\mathit{ff}$ terms in the Lagrangian are suppressed and start at  linear order in curvature $\mathcal{O}(m^2)$.  This result is consistent with the vanishing of the Maxwell
 term for $v=0$ in the decoupling limit uncovered in Ref.~\cite{deRham:2010tw}.

The $f\aM$ terms follow a similar pattern.
For  $\dot T \neq x_0 + \mathcal{O}(m)$, the coefficients $K_2, K_4$ start in the linear order in $m$, while the other two coefficients $K_1, K_3$ are suppressed by an additional power of $m$ and start at $\mathcal{O}(m^2)$. If $\dot T = x_0 + \mathcal{O}(m)$ then all these coefficients start at the order $\mathcal{O}(m^2)$ and this is thus also order at which we recover the dynamics of the \stucky\ perturbations.

There are also time derivative terms from the \stucky-metric contributions.   In fact there are two terms with time derivatives on $\aM_t$, $h_{\theta\theta} \aM_{t;t}$ and $h_{\phi\phi} \aM_{t;t}$ which might seem problematic for the non-dynamical nature of $\aM_t$. 
However, as argued in Ref.~\cite{Mirbabayi:2011aa}, these do {not change the} dynamics and hence the
reappearance of the Boulware-Deser ghost  because the derivatives can be moved to $h_{\theta\theta}, h_{\phi\phi}$ by integration by parts.  This integration by parts leaves $\aM_t$ manifestly nondynamical, while not disturbing the non-dynamical nature of  $h_{0\mu}$ for
imposing constraints.

It turns out that in the flat patch approximation $\aM h$ coupling gives kinetic mixing terms to the spatial \stucky\ $\aM_r, \aM_\theta, \aM_\phi$ at  $\mathcal{O}(m)$ for the case $\dot T \neq x_0 + \mathcal{O}(m)$, while in the case without the vector in the background $\dot T = x_0 + \mathcal{O}(m^2)$ these kinetic terms start at $\mathcal{O}(m^2)$.
The metric-metric Lagrangian has kinetic terms from only the usual Einstein-Hilbert Lagrangian.   We thus conclude that as expected the full Lagrangian generally has kinetic
terms for the 3 spatial \stucky\ fields and the usual 2 tensor modes for a total of 5 modes.
In no case are there \stucky\ kinetic terms
at ${\cal O}(m^0)$ consistent with \S \ref{sec:decoupling} and Ref.~\cite{D'Amico:2012pi}.
For special cases they may begin at ${\cal O}(m^2)$ or be absent entirely.

\subsection{Example}
\label{sec:koyamaform}

To make these considerations concrete, we return here to the specific solutions considered in
\S \ref{sec:decoupling}. Recall that these solutions are for the
 $\alpha_3 = \alpha_4 = 0$ case where $P_1'(x_0)=-4$.

For these background solutions we have 
\ba
	\bar \OurG_{tt} &=& \(1 - \frac{1}{C}\) - \frac{m^2\(r^2(1+C^2) + t^2 v^2 C^2\)}{8 C(1+C)},\nonumber\\
	\bar \OurG_{tr} &=& -\frac{ m r v}{2(1+C)} - \frac{ m^2 r(t v^2 C + 2 r \scalar)}{8(1+C)},\nonumber\\
	\bar \OurG_{rr} &=& -\frac{m^2 r^2}{4C(1+C)} ,
\ea
plus terms which are higher order in graviton mass. The remaining components are given by the general formulae as described in the previous section.   Note that this explicit 
form is consistent with the general considerations of Eq.~(\ref{bargammaexp}) and
(\ref{bargammav0})
for the $v\ne 0$ and $v=0$ cases respectively.

Using the results of the previous section, we can then write down the \stucky-\stucky\ quadratic Lagrangian as
\ba
\label{KoyamaLSS}
	&& \frac{\sqrt{-\bar g}}{ \sin \theta} L_{SS}= - \frac{(4 v^2 C^2 + m^2 r^2)}{64 C(1+C)}\(f_{r\phi}^2\csc^2 \theta + f_{r\theta}^2\)
	\nonumber\\
	&&\quad -
	\frac{ v^2 C\csc^2 \theta}{16 (1+C)r^2} f_{\theta\phi}^2 
	 - \frac{m^2 r^2}{64 C(1+C)}\(f_{t\theta}^2 + f_{t\phi}^2\csc^2 \theta\)
	\nonumber \\
	&&\quad + {mr} \frac{4  v + m (t C v^2 +2  r \scalar)}{64(1+C)}
	 \(f_{r\theta}f_{t\theta} + f_{r\phi}f_{t\phi} \csc^2 \theta\)
	\nonumber \\
	&&\quad - \frac{m^2 r(2+C)}{16(1+C)}\(\aM_\theta f_{r\theta} + \aM_{\phi} f_{r\phi} \csc^2 \theta\)
	\nonumber \\
	&&\quad - m \frac{4 C v + 3 C m r \scalar}{16(1+C)}\(\aM_\theta f_{\theta t} + \aM_\phi f_{\phi t} \csc^2 \theta\)
	\nonumber \\
	&&\quad -mr^2 \frac{ m r \aM_t-(2 C v + C m r \scalar) \aM_r  }{8(1+C)}f_{tr}
	\nonumber \\
	&&\quad - \frac{m^2r^2}{16(1+C)}[ (1+2C) \aM_r^2 - 3 C \aM_t^2 ]
	\nonumber \\
	&&\quad - \frac{m^2 (1+2C)}{16(1+C)} \(\aM_\theta^2 + \aM_\phi^2 \csc^2 \theta\). 
\ea
Even if we ignore kinetic mixing with the metric, a Hamiltonian analysis shows that
the \stucky-\stucky\ Lagrangian itself propagates  both transverse modes $\aM_\theta, \aM_\phi$ and the longitudinal mode $\aM_r$, giving three dynamical degrees of freedom. This Hamiltonian is unbounded with respect to the spherically symmetric perturbations $\aM_\theta = \aM_\phi = 0$.   This is related to the unboundedness of $a_r$, $a_t$ from 
\S \ref{sec:decoupling} since
\ba
	\aM_t &=& a_t(t,r) + m r \frac{v C}{2(1+C)} a_r(t,r) + \mathcal{O}(m^2),\nonumber\\
	\aM_r &=& a_r(t,r) + m r \frac{v C}{2(1+C)} a_t(t,r) + \mathcal{O}(m^2).
\ea
Note that the $v=0$ case is also special in that terms from the tetrad alignment do not
appear until $\mathcal{O}(m^2)$. 

The \stucky-\stucky\ Maxwell terms follow the general behavior pointed out in the previous section. For the case of no vector in the background $v=0$ and $C = 1$, all terms start at  most at $\mathcal{O}(m^2)$, while in the other cases spatial derivative terms  $f_{r\theta}^2, f_{r\phi}^2$ and $ f_{\theta\phi}^2$ start at $\mathcal{O}(m^0)$ and
$f_{r\theta} f_{t\theta}, f_{r\phi} f_{t\phi}, \aM_r f_{tr}, \aM_{\theta} f_{\theta t}, \aM_{\phi}f_{\phi t}$ start at $\mathcal{O}(m)$. 

Focusing on the terms which appear before $\mathcal{O}(m^2)$, only $f_{r\theta} f_{t\theta}, f_{r\phi}f_{t\phi}$ can provide any dynamics in $\mathcal{L}_{SS}$ as the remaining time derivatives can be integrated out. However, as the Hamiltonian analysis shows, the $\mathcal{O}(m)$ \stucky-\stucky\ Lagrangian does not propagate all three modes and we have to go to $\mathcal{O}(m^2)$ if we want to capture the correct dynamics with $\mathcal{L}_{SS}$ only. This once more stresses the importance of retaining all $\mathcal{O}(m^2)$ terms in the Lagrangian to correctly describe the dynamics of the system. 


\section{Discussion}
\label{sec:discussion}

We have provided a complete and covariant treatment for the quadratic Lagrangian of all of the degrees of freedom
of massive gravity with a fixed flat fiducial metric around any isotropic self-accelerating background for any set of massive gravity parameters.
We find that for generic cases 3 out of 4 \stucky\ degrees of freedom propagate
in addition to the usual 2 tensor degrees of freedom of general relativity.   The complete
kinetic structure typically is only revealed at ${\cal O}(m^2)$ or equivalently 
curvature terms in a locally flat expansion.   

These results resolve a number of apparent discrepancies in the literature.   
The kinetic terms for all additional degrees of freedom vanish in the leading order, Minkowski term in the locally flat approximation and are only fully established at the order of
curvature corrections, omitted in the analysis of Ref.~\cite{D'Amico:2012pi}. This result  differs from  the usual Minkowski
decoupling limit because on the self-accelerating branch of solutions there is always
a non-trivial background \stucky\ field.   Because the Minkowski scaling limit is not justified around self-accelerated solutions, analyses that are based on it can be misinterpreted.  It is important
to distinguish between an imposed scaling of parameters with the graviton mass 
and a true decoupling limit
where degrees of freedom are both preserved and decoupled.  It is also important to
note that \stucky\ fields restore gauge invariance and the redundancy that exists
because of their introduction should be fixed as a gauge freedom.   Together they 
explain the results of Ref.~\cite{Koyama:2011wx}.  Finally the  case
of open universe solutions where the spacetime and fiducial metrics  are simultaneously diagonal, homogeneous and isotropic
 is extremely special and propagate no degrees
of freedom about the exact solution \cite{Gumrukcuoglu:2011ew}.

The covariant quadratic Lagrangian exhibits several notable and potentially problematic
features.   Spatial derivatives of the degrees of freedom can appear at a lower order
than temporal derivatives.   Relatedly,  as  shown in Ref.~\cite{Wyman:2012iw},
anisotropic stresses can dominate the stress energy tensor of fluctuations.  
We leave a full stability analysis of  the joint \stucky,  metric, and matter system for a future work.

\smallskip{\em Acknowledgments.--}  
We thank P.\ Adshead, G.\ D'Amico, P.\ Gratia, A.\ Joyce, K.\ Koyama, M. Mirbabayi, and L.T.\ Wang for useful discussions.
This work was  supported
 by U.S.~Dept.\ of Energy
 contract DE-FG02-13ER41958
 and the
 Kavli Institute for Cosmological Physics at the University of
 Chicago through grants NSF PHY-0114422 and NSF PHY-0551142.

\bibliography{MassiveV}

\end{document}